\documentstyle[12pt]{article}
\newcommand{\junk}[1]{}

\newcommand{\Z}{Z\!\!\!Z}
\newcommand{\beq}{\begin{equation}}
\newcommand{\eeq}{\end{equation}}
\newcommand{\bea}{\begin{eqnarray}}
\newcommand{\eea}{\end{eqnarray}}
\newcommand{\bd}{\begin{displaymath}}
\newcommand{\ed}{\end{displaymath}}
\newcommand{\nn}{\nonumber}

\newcommand{\N}{{\cal N}}

\newcommand{\g}{{g_{\rm st}}}
\newcommand{\G}{{\cal G}}
\renewcommand{\H}{{\cal H}}

\newcommand{\im}{{\rm Im}~}
\newcommand{\h}{{\tilde h}}
\renewcommand{\g}{{\tilde g}}

\newcommand{\m}{{\tilde{ m}}}
\newcommand{\tsigma}{{\tilde \sigma}}
\newcommand{\dd}{{\rm d}}
\newcommand{\D}{{\cal D}}
\newcommand{\R}{{{\rm l\!R}}}

\begin{document}
%%%%%%%%%%%%%%%%%%%%%%%%%%%%%%%%%%%%%%%%%%%%%%%%%%%%%%%%%%%%%%%%%%%%%%%%%%%

\thispagestyle{empty} 
{\hfill {NBI-HE-98-15}}\\[10mm]
\centerline{{\Large{\bf
$T$-Duality in Lattice Regularized $\sigma$-Models}}}
\vskip5mm 
\centerline{\bf Sebastian Jaimungal\footnote{email: jaimung@nbi.dk, jaimung@physics.ubc.ca \\ This work is supported in part by NSERC of Canada.}} \vskip 5mm
{\it
\centerline{The Niels Bohr Institute}
\centerline{Blegdamsvej 17}
\centerline{DK-2100 Copenhagen $\emptyset$, Denmark.}
\vskip 2mm
\centerline{and} \vskip 2mm
\centerline{Department of Physics and Astronomy}
\centerline{University of British Columbia} 
\centerline{6224 Agricultural Road} 
\centerline{Vancouver, British Columbia V6T 1Z1, Canada.}
}
\vskip 3mm

\begin{abstract}
\noindent
It is shown that when the underlying sigma model of bosonic string
theory is written in terms of single-valued fields, which live in the
covering space of the target space, Abelian $T$-duality survives
lattice regularization of the world-sheet. The projection onto the
target-space is implemented through a sum over cohomology, which bears
resemblance to summing over topological sectors in Yang-Mills theories. In
particular, the case of string theory on a circle is shown to be
explicitly self-dual in the lattice regulated model and automatically
forbids vortex excitations which would otherwise destroy the
duality. For other target spaces a generalized notion of $T$-duality
is observed in which the target space and the cohomology coefficient
group are interchanged under duality. Specific examples show that the
fundamental group of the target space may not be preserved in the
$T$-dual theory.  Generalized models which exhibit $T$-duality behaviour,
with dynamical variables that live on the $k$-dimensional cells of
$(p+1)$-dimensional world-volumes, are also constructed. These models
correspond to gauge theories, and higher-dimensional analogues, in
which one sums over various topological sectors of the theory.
\end{abstract} 

\vskip 4mm


\newpage\setcounter{page}{1}

\setcounter{equation}{0} 
%
%
% Introduction
%
%
Target space duality\cite{Tdual} is a symmetry of string theory which
maps models defined on classically distinct target manifolds into one
another. This is a rather surprising result when observed from the
point of view of the target space. However, it has been known
for some time now that the underlying principle of $T$-duality is
intimately connected with the Hodge duality of forms on the
world-sheet and is manifest in the sigma model that defines the theory
\cite{MoOv89, Du90}.  An interesting and useful question to ask is
whether this duality survives lattice regularization of the
world-sheet. The authors of \cite{Si96} determined the potential, for
a $D=0$ matrix model, which preserved $T$-duality at the level of
Feynman graphs.  The question of whether this duality survives when
spins are included on the sites of the graph was first studied by
\cite{GrKl90} where a string partition function for a discretetized
world-sheet with target space $S^1$ was formulated. They noticed that
when the Hodge duality was applied on the lattice, which amounts to
performing a Kramers-Wannier $S$-duality transformation \cite{KrWa41},
the $T$-duality of the continuum model was lost due to the existence
of vortices, and the model undergoes a Kosterlitz-Thouless phase
transition at a critical radius of the target space.  This loss of
$T$-duality was seen as a lattice artifact and was solved by altering
the string partition function to forbid all vortex configurations. An
ansatz which implemented this restriction was inserted by hand and the
model regained its self-dual nature. In this letter we show that if one 
defines the partition function in terms of single-valued fields that
are elements of the cover of the target space, while a sum over
harmonic forms induces the projection down to the target space, then a 
straightforward lattice regularization of the model immediately
implements such a constraint. The
sum over harmonic forms can be thought of as a sum over large gauge
transformations and is analogous to the schemes implemented in
refs. \cite{ShVa88, KoSh97, HaZh97} in which sums over the different
theta sectors of (Super)-Yang-Mills theory was introduced in order to
produce the correct $2\pi$ periodicity of the various correlators.

The idea of lifting from the target space, $S^1$, to the covering
space, $\R$, is quite similar to what one does when quantizing on the
circle \cite{Sc81}. This general strategy will be applied to
non-linear sigma-models with target space $\G/\H$ (throughout this
letter $\G$ and $\H$ are Abelian groups) where the world-sheet has
been regulated by a lattice. The spins will be taken to be elements of
the natural cover of $\G/\H$, which is $\G$, and the projection onto
the target space will be implemented through a sum over $\H$-valued
cohomology. On the lattice a generalized idea of duality is observed
in which the target space and the coefficient group of the cohomology
are interchanged.  This is the analogue of the momentum and winding
modes being interchanged under duality in the continuum theory. We
determine choices of $\G$ and $\H$ which renders the model explicitly
self-dual. In addition to the usual self-dual $S^1$ target space, we
identity the target space $\Z_N$ with cover $\Z_{N^2}$ as new
self-dual models. These models are the discrete versions of the
circular target space.  We also construct generalized models in which
a $\G/\H$-valued ``spin'' lives on the $(k-1)$-dimensional cells of a
$(p+1)$-dimensional triangulated world-volume. These models have the
interpretation of a $p$-brane on which a $\G/\H$-valued $(k-1)$-form
field is defined. If, however, the world-volume is viewed as
space-time, then the models correspond to modified theories of spins,
gauge fields, antisymmetric tensors fields, etc.., where one sums over
various topological sectors of the theory.
%
%
% Continuum..
%
%
%

We begin with a discussion of $T$-duality in the continuum (for a 
review see for instance \cite{AlAlGLo95, GiPoRa94}), where the 
anti-symmetric tensor field is absent.  The sigma model action
is given by, \beq S =
\frac{1}{\sqrt{\alpha '}}\int_{\Sigma} d^2 \sigma \; G_{\mu\nu}\;
\partial X^\mu {\overline \partial} X^\nu \label{PolyaModel} \eeq here
$\Sigma$ is a fixed, but arbitrary, orientable two-dimensional
world-sheet of genus $g$, $G^{\mu\nu}$ is the target space metric and
we have trivialized the world-sheet metric. Consider the case in which
the target space is $S^1$ and write $X^0\equiv\theta$. The mode
expansion of the $\theta$ co-ordinate contains, in addition to the
vibrational modes, winding modes corresponding to the string wrapping
around the target space. Hence, $\theta$ can be multi-valued,
consequently as one moves along a non-contractable loop of the
world-sheet $\theta$ can pick up an extra factor of $2\pi\times {\rm
integer}$, i.e.,
\beq
\oint_{\gamma_a} \dd\theta \in 2\pi \Z \label{multi}
\eeq
where $\{\gamma_a:\,a=1,\dots,2g\}$ are the canonical set of cycles
which generate the first singular homology group of $\Sigma$. Defining
the partition function in terms of multi-valued fields is undesirable
both from a pedagogical point of view, and since multi-valued fields
do not exist in lattice regularization.  Lifting $\theta$ to the
covering space of the circle, i.e. the real line, allows a natural
decomposition into a smooth single-valued function and an element of
the integer cohomology of the world-sheet: $\dd\theta \to \dd\theta +
2\pi h$. The cohomology elements are
the analogues of the winding modes in the mode expansion. It is
important to realize that this ansatz for introducing single-valued
fields is quite general.  If the $\theta$ co-ordinate takes values in
$\G$ and the cohomology in $\H$, then the target space is the quotient
space $\G /\H$, which has $\G$ as a natural cover.  To make contact
with the work of \cite{GrKl90} we first develop the case when $\G =
\R$ and $\H=\Z$, in the latter part of this letter we will introduce
the general models on the lattice. The new partition function (for a
fixed surface $\Sigma$) is then written as,
\beq Z = \sum_{h\in H^1(\Sigma,\Z)} \int
\D \theta \exp \left\{ - \frac{1}{\sqrt{\alpha'}} \int_\Sigma G^{00}
(\dd\theta + 2\pi h) \wedge *(\dd\theta + 2\pi h)\right\}
\label{ContModel} \eeq 
This will be the defining continuum theory, and all models introduced
in this letter are straightforward lattice regularizations of this
partition function and simple modifications of the coefficient
groups. Notice that here there is an explicit sum over several
partition functions each defined on the cover of the target space and
from the outset multi-valued fields are absent. Although it is a
simple re-writing of the model, we will see that its lattice
regularization leads to an explicitly self-dual theory without the
insertion of any extra constraints.

We now demonstrate that (\ref{ContModel}) is explicitly self-dual.
The strategy is a familiar one, first introduce a one-form $V$ which
satisfies a Bianchi constraint plus holonomy constraints, 
\beq
Z = \int
\D V \;\delta\left(*\dd V\right) \left(
\prod_{a=1}^{2g} \delta_{2\pi} \left( \oint_{\gamma_a} V \right)
\right) \exp\left\{-\frac{1} {\sqrt{\alpha'}} \int_\Sigma G^{00} (V \wedge
*V)\right\} \label{ContConstr}
\eeq
Notice that there are two distinct types of delta-functions here, the
first constraint implies that $V$ is the sum of an exact form plus
cohomology elements with real coefficients, while the second periodic
constraint forces the coefficients of the cohomology to be elements of
$2\pi\Z$.  Solving the constraints on $V$ leads back to the original
model (\ref{ContModel}). Alternatively one can introduce Lagrange
multipliers to implement the constraints, $$ Z= \int
\D V \; \D{\tilde \theta} \sum_{{\tilde h}\in H^1(\Sigma, \Z)} 
\exp\left\{-\frac{1} {\sqrt{\alpha'}} \int_\Sigma G^{00} (V \wedge *V) 
+ i\; \dd V \wedge \tilde \theta + i\; V \wedge {\tilde h} \right\} 
$$
To represent the holonomy constraints we have used: $\oint_{\gamma_a} V =
\int_{\Sigma} V \wedge h^a$ where $\{h^a\}$ are the canonical set of
cohomology elements dual to the cycles $\{\gamma_a\}$:
$\oint_{\gamma_a} h^b = \delta_a^b$.  The one-form $V$ now appears
only quadratically in the action, and it can be eliminated via its
equations of motion to give the dual partition function\footnote{The
determinant factor only serves to shift the dilaton which we ignore
here.}, $$ Z = \sum_{{\tilde h}\in H^1(\Sigma, \Z)} \int \D {\tilde
\theta}
\exp \left\{ - \sqrt{\alpha'} \int_\Sigma \frac 1{4G^{00}} ~(\dd\tilde
\theta + {\tilde h}) \wedge *(\dd\tilde\theta + {\tilde h}) \right\}
$$ Here the replacement of the old fields with the Lagrange
multiplier fields is the analogue of interchanging the winding and
momentum modes in the mode expansion of the string co-ordinates.  Of
course, in addition, the target space metric transformed as $G^{00}
\leftrightarrow \alpha'/4G^{00}$.  
%
%
% Gross-Klebanov model..
%
%
%

It is instructive to demonstrate that a lattice regularization of
(\ref{ContModel}) eliminates vortex configurations and leads directly
to the constrained model in \cite{GrKl90}. In order to
express the results in a manner which is easily generalized, we
introduce some notations of simplicial homology (see for example
\cite{Sp66}).  Let $\Sigma$ be a triangulation of a smooth manifold
 and $\{c_k^{(i)}\}$ be the generators of
the chain complex $(C_*(\Sigma,\G),\partial)$ with Abelian coefficient
group $\G$ (group multiplication is written additively) and boundary
operator $\partial$ defined by the incidence numbers~$[\cdot, \cdot]$,
$$\partial_k c_k^{(i)} = \sum_{j=1}^{\N_k} [ c_k^{(i)} : c_{k-1}^{(j)}
] \,c_{k-1}^{(j)} $$ Here $\N_k$ denotes the number of $k$-cells in
the lattice $\Sigma$ and $[c_k^{(i)}, c_{k-1}^{(j)}]$ is $\pm 1$ if
the cell $(j)$ is contained in the cell $(i)$ and zero otherwise.
There exists a natural inner product between the generators, $$
\langle c_k^{(i)} , c_l^{(j)} \rangle = \delta_{k,l} \delta^{i,j}
$$ 
which acts linearly on elements of the chain complex. This inner product
induces the operation of the co-boundary operator, $\delta$,
$$
\langle \partial g , h \rangle = \langle g , \delta h \rangle 
$$ where $g$ and $h$ are arbitrary chains (elements of $C_*(\Sigma,
\G)$). The simplicial homology and cohomology groups are then given by, $$
H_k(\Sigma, \G) = \ker \partial_{k}/\im \partial_{k+1} \quad, \quad
H^k(\Sigma, \G) = { \ker \delta_{k}}/ {\im \delta_{k-1}} $$ A
straightforward lattice regularization of (\ref{ContModel}) can then
be written as,
\beq
Z = \sum_{h\in H^1(\Sigma, \Z)} \;\sum_{\sigma\in C_0(\Sigma, \R)}\;
\prod_{l=1}^{\N_1} B\left( \left\langle (\delta\sigma + 2\pi h) , c_1^{(l)} 
\right\rangle\right) \label{LatticeRegModel}
\eeq Here the spins $\sigma$ are the analogue of $\theta$ in the
continuum and the Boltzmann weight is defined by $B(g) \equiv \exp\{ -
(\alpha')^{-1/2} G^{00} g\}$. It is possible to introduce a
real-valued one-chain in place of the spins and cohomology, much like
introducing the real-valued one-form $V$ in the continuum. This leads
to the following representation,
\beq Z = \sum_{v\in C_1(\Sigma,\R)} \;
\left(\prod_{l=1}^{\N_1}
\delta_{\R} \left( \langle \delta v ,c_1^{(l)} \rangle \right) \right)\;
\left( \prod_{a=1}^{2g} \delta_{U(1)}
\left( \langle v, h_a \rangle \right) \right) \; 
\prod_{l=1}^{\N_1} B\left( \left\langle v, c_1^{(l)} \right\rangle\right)
\label{GrKlConRegMod}
\eeq 
where $h_a \equiv \sum_{l\in\gamma_a} c_1^{(l)}$ are the generators of
the first homology group.  In the above $\delta_{\G}$ represents a
$\G$ invariant delta function. This form of the partition function is
the lattice analogue of (\ref{ContConstr}), the first constraints are
the Bianchi constraints forcing $v$ to be the sum of a co-exact chain
and a cohomology element both with real coefficients; while the second
constraints forces the coefficients of the cohomology to be elements
of $2\pi\Z$. Consequently, solving the constraints on $v$ reproduces
the original model much like in the continuum.  There exists a
slightly different representation of the model which makes direct
contact with the work of \cite{GrKl90}. This involves decomposing
$\sigma$ into an integer valued chain, $\tsigma$,(integer valued
fields pose no problem on the lattice) and a $U(1)$ valued chain,
$\theta$.  Decomposing $\sigma$ in this manner allows one to introduce
an integer-valued one-chain in place of $\tsigma$ and $h$ while
leaving $\theta$ untransformed. In this representation there are
dynamical variables on the sites and links of the lattice. The partition 
function in this decomposition is given by the following,
\bea Z &=&\sum_{h\in H^1(\Sigma,
\Z)}\;\sum_{\theta\in C_0(\Sigma,U(1))}\;
\sum_{{\tsigma}\in C_0(\Sigma, \Z)}\; \prod_{l=1}^{\N_1} B\left(
\left\langle (\delta \theta + 2\pi(\delta {\tilde\sigma} + h)) ,
c_1^{(l)} \right\rangle\right) \nn\\ &=& \sum_{\theta\in
C_0(\Sigma,U(1))}\;\sum_{v\in C_1(\Sigma,\Z)} \left(
\prod_{l=1}^{\N_1} \delta_{\Z} \left( \langle \delta
v,c_1^{(l)}\rangle \right) \right) \prod_{l=1}^{\N_1} B\left(
\left\langle (\delta \theta + 2\pi v), c_1^{(l)} \right\rangle\right)
\label{GrKlmodel} \eea  
Notice that here there are no lattice analogues of the holonomy
constraints as in (\ref{GrKlConRegMod}). The advantage in writing the
model in its present form is that the spin variables take values in
the target space itself rather than its covering space. This is a
desirable prescription, however, the decomposition which leads to this
model is not well-defined in the continuum as integer valued fields
are problematic, and is thus only valid on the lattice. In this
representation the winding modes are implemented through the action of
the link-valued objects $v$. As one moves around elementary plaquettes
the winding number, given by $\sum_{l\in p} v_l$, must vanish due to
the Bianchi constraint, while along the canonical cycles there is no
restriction on the winding number. This is precisely the constraint
that the authors of \cite{GrKl90} inserted into the discrete version
of (\ref{PolyaModel}), which they wrote as an $X-Y$ model on $\Sigma$,
in order to suppress vortex configurations.  It is, however, clear
that these constraints follow directly from a lattice regularization
of (\ref{ContModel}) and there is no need to insert it by hand. This
feature is a direct consequence of writing the continuum variables as
single valued fields in the covering space of the circle and then
projecting onto the target space through a sum over cohomology
elements.

Let us now perform the duality transformation on this model. On the
lattice it is easiest to perform the transformations directly on
(\ref{LatticeRegModel}) rather than on (\ref{GrKlConRegMod}) or
(\ref{GrKlmodel}).  Inserting a character expansion (in this case a
character expansion amounts to a Fourier transformation) of the
Boltzmann weights in the partition function introduces a
representation on every link, encoding this information into a
one-chain, denoted by $r$, one finds, 
\bea Z &=& \sum_{h\in H^1(\Sigma, \Z)}
\;\sum_{\sigma\in C_0(\Sigma, \R)}\;
\prod_{l=1}^{\N_1} \sum_{r_l\in\R} \;b(r_l) \;\chi_{r_l} \left(
\left\langle (\delta\sigma + 2\pi h) , c_1^{(l)} \right\rangle\right) \nn\\
&=& \sum_{r\in C_1(\Sigma, \R)} \; \prod_{l=1}^{\N_1} b \left( \langle r,
c_1^{(l)} \rangle \right) \; \sum_{h\in H^1(\Sigma, \Z)}
\;\sum_{\sigma\in C_0(\Sigma, \R)}\;
\prod_{l=1}^{\N_1} \chi_{\langle r , c_1^{(l)} \rangle} \left(
\left\langle (\delta\sigma + 2\pi h) , c_1^{(l)} \right\rangle\right)
\label{CharExpGK}
\eea here the character coefficients of the Boltzmann weights are
given by $b(r) \equiv \sum_{g\in\R} {\overline \chi}_{r} ( g) B(g)$
(throughout we ignore overall constants which can easily be
restored). The characters satisfy simple factorization properties,
$$ \chi_{r_1}(g) \chi_{r_2}(g) = \chi_{r_1+r_2} (g), \qquad
\chi_{r}(a g) = \chi_{a r}(g) $$ which can be used
to re-arrange the sums over $h$ and $\sigma$ in (\ref{CharExpGK}) into
the form, 
$$ \sum_{h,\sigma} \dots \; =
\left(\prod_{i=1}^{\N_0} \sum_{\sigma_i\in \R} \chi_{\langle
\partial r, c_0^{(i)} \rangle} \left( \sigma_i \right)\right)\!\!
\left(\prod_{a=1}^{2g} \sum_{m_a\in\Z}\chi_{\langle r, h^a\rangle}
\left ( 2\pi m_a \right)\right)
%\nn\\ 
= \delta_{\R} \left( \partial r \right)
\prod_{a=1}^{2g}\delta_{U(1)} \left(2\pi \langle r, h^a \rangle \right)
$$
The orthogonality of the characters was used to obtain the last
equality and $h^a$ is the generator of the cohomology dual to homology
generator $h_a$: $\langle h_a, h^b \rangle = \delta_a^b$. The first
constraint forces $r$ to be closed and is therefore a sum of an exact
chain and an element of the homology group both with real coefficients: $r=
\partial \tsigma + \h$; while the
second constraint forces the coefficients of the homology to be integers. 
Inserting this solution of the constraints into (\ref{CharExpGK}) yields,
$$
Z = \sum_{{\tilde h} \in
H_1(\Sigma, \Z) } \; \sum_{{\tsigma}
\in C_0(\Sigma,\R) } \;\prod_{l=1}^{\N_1} b \left( \left\langle
\partial {\tsigma} + {\tilde h} , c_1^{(l)} \right\rangle \right)
$$
Interpreting the generators,
$\{c_k^{(i)}\}$, of the chain complex on the dual lattice transforms the
boundary operator to a co-boundary operator and homology to cohomology,
the dual partition function then reads, $$ Z = \sum_{{\tilde h} \in
H^1(\Sigma^*, \Z) } \; \sum_{{\tsigma}
\in C_0(\Sigma^*,\R) } \;\prod_{l=1}^{\N^*_1} b \left( \left\langle
\delta {\tsigma} + {\tilde h} , c_1^{*(l)} \right\rangle \right) 
$$ where the starred objects are on the dual lattice. This is clearly
equivalent to the original model (\ref{LatticeRegModel}) with the
Boltzmann weights being replaced by their character coefficients, 
$$ B(g) = \exp\left \{ - \frac{R^2}{\sqrt{\alpha'}} g^2 \right \}
\;,\qquad b({\tilde g}) = \sqrt{\frac{\pi\alpha'^{\frac12}}{R^2}}
\;\exp\left \{ - \frac{\sqrt{\alpha'}}{4R^2} {\tilde g}^2\right\} $$
Thus we recover the $T$-duality transformation of the continuum model,
with the lattice being replaced by the dual lattice and
$R\leftrightarrow\sqrt{\alpha'}/2R$. We have demonstrated that writing
the continuum model in terms of the covering space of the circle and
performing a straightforward lattice regularization, leads to an
automatic suppression of vortex configurations and to an explicitly
self-dual model.
%
%
%
% Generalized Models: G/H
%
%

We would like to generalize the model to include target spaces which
are the quotient of two arbitrary Abelian groups $\G/\H$ in which $\H$
acts freely on $\G$. With the world-sheet regulated by a lattice the
model is a trivial extension of (\ref{LatticeRegModel}) and is written
as, \beq Z = \sum_{h\in H^1(\Sigma, \H)} \;\sum_{\sigma\in
C_0(\Sigma,\G)}\; \prod_{l=1}^{\N_1} B\left( \left\langle
\left(\delta\sigma + h \right) , c_1^{(l)} \right\rangle\right) 
\label{GHmodel}\eeq
Here elements of $\H$ are written is such a way so that addition in
the Boltzmann weight is well-defined. For example, if $\G=U(1)$ and
$\H=\Z_N$, then the argument of the Boltzmann weight should be:
$\delta\sigma + (2\pi/N)h$.  These models are very similar to the ones
considered in \cite{GaJaSe98} where the authors consider a $\G$-valued
spin model and introduced a sum over a subset of the
generators of the cohomology in order to generate
self-dual models. The difference here is that the sum extends over the
entire cohomology and its coefficient group differs from the spins
coefficient group, while in \cite{GaJaSe98} they were taken to be
identical.

Performing the dual transformations on (\ref{GHmodel}) is a simple
generalization of the previous calculation and in lieu of repeating the
steps we mention the relevant points.  A character expansion of the
Boltzmann weights is carried out and the one-chain $r$ carries an element
of $\G^*$ (the group of irreducible representations of $\G$, for
Abelian groups $\G^*$ inherits the groups Abelian structure) on every
link. The factorization properties of the characters allows the sum
over $h$ and $\sigma$ to be performed and constrains $\partial r$ to
vanish in $\G^*$ and $\langle r, h^a
\rangle$ to vanish in $\H^*$. The first constraint forces $r$ to be a
sum of an exact chain and an element of the homology group with
$\G^*$ coefficients, while the second set of constraints forces the
coefficient group of the homology to be $\G^*/\H^*$. Interpreting the
objects on the dual lattice leads to the dual model,
\beq
Z =
\sum_{\h\in H^1(\Sigma, \G^*/\H^*)}
\;\sum_{\tsigma\in C_0(\Sigma,\G^*)}\;
\prod_{l=1}^{\N_1^*} b\left( \left\langle \left(\delta\tsigma + \h
\right) , c_1^{*(l)} \right\rangle\right)
\eeq

The effects of the duality transformation on the various coefficient
groups and the target space are shown in Table \ref{GrTrans}. This
table illustrates an interesting generalized version of
$T$-duality. Unfortunately, if either the original or dual spin
variables take values in a discrete group, then these models can only
be defined on the lattice . This is simply because a continuum theory
cannot have discrete valued fields. Nevertheless, the lattice models
are perfectly well-defined, and we should investigate what the duality
implies.  It is interesting to identify the groups which lead to
explicitly self-dual models. Certainly a necessary condition is that
$\G \cong\G^*$ (spin models on $2$-d infinite lattices also have this
self-dual restriction).  In that case, under duality the coefficient
group of the cohomology and the target space are interchanged and then
replaced by their dual group. This is the analogue of the
interchanging of the momentum and winding modes and $R\leftrightarrow
\sqrt{\alpha'}/R$ in the mode expansion of the string co-ordinates. It
is also the analogue of the cohomology group being replaced by the
Lagrange multipliers which implement the holonomy constraints in the
path integral.
\begin{table}
\begin{center}
\begin{tabular}{|r|c|c|c|} \hline
      & Spin Variable &Cohomology Coefficient & Target Space\\ \hline
%      &               & Coefficient & \\ \hline
Original Model  & $\G$           & $\H$                        & $\G/\H$ \\
\hline
Dual Model     & $\G^*$         & $\G^*/\H^*$                 & $\H^*$ \\
\hline
\end{tabular}
\caption{Transformations of the various groups under duality.}\label{GrTrans}
\end{center}
\end{table}

Consider the model defined in eq. (\ref{LatticeRegModel}). We will
obtain its dual using Table \ref{GrTrans}. In that case $\G = \R$ and
$\H = 2\pi R\Z$ so that the target space is $\R/2\pi R\Z\cong S_R^1$
where the subscript identifies the radius of the circle. The dual
model has $\G' = \R^*\cong \R$, $\H' =\G'/2\pi R\Z^*
\cong R^{-1} \Z$ and target space $\G'/\H' \cong S_{(2\pi
R)^{-1}}^1$. We have thus recovered the earlier result that the duality
transformation only serves to invert the radius of the target space.
Notice that it is straightforward to write down the result for a
toroidal target space $T^n = S^1 \times \dots \times S^1$. In that
case one chooses $\G = \R \oplus \dots \oplus \R$ and $\H = 2\pi R_1
\Z \oplus
\dots\oplus 2\pi R_n \Z$ the target space is obviously the n-tori with
compactification radii $R_i$ in the $i$-th direction.  The dual model
leaves $\G$ invariant as $\G'=\G^*\cong\G$ while $\H' = \G'/\H^* =
R_1^{-1} \Z \oplus \dots \oplus R_n^{-1} \Z$ and the dual target space is
an n-tori with radii $(2\pi R_i)^{-1}$ in the $i$-th direction.
This case corresponds to taking the target space metric to be
diagonal. Of course it is possible to consider metrics with off diagonal
elements. To incorporate this into our formalism the
Boltzmann weights should be defined as follows,
$$
B((g_1,\dots,g_n)) = \exp \left\{ - G^{ij} g_i g_j \right\} ,
\qquad 
b((\g_1,\dots,\g_n)) = \frac{1}{4\pi\sqrt{G}}
\exp \left\{ -\frac 1 4 (G^{-1})^{ij} \g_i \g_j\right\}
$$ here the $n$-tuple $(g_1,\dots,g_n)$ and $(\g_1,\dots,\g_n)$
represent elements of $\G = \R\oplus\dots\oplus\R$ and $\G'\cong\G$
respectively. Also, choose $\H = 2\pi\Z\oplus\dots\oplus 2\pi\Z$ so
that the metric information is contained solely in the Boltzmann
weight. This demonstrates that under duality the target space metric
is replaced by its inverse and reduces to one of the Buscher
formulae\cite{Bu87} in the case of vanishing torsion.

The toroidal compactifications are the simplest example of a self-dual
model. There are other groups which satisfy the necessary condition
$\G\cong\G^*$ namely the cyclic groups $\Z_P$. For this choice of $\G$
the coefficient group of the cohomology is forced to be cyclic as
well, $\H=\Z_N$ where $N$ is a factor of $P$ (let $P=NM$). This is
necessary so that the action of $\H$ on $\G$ (identifying elements of
$\G$ which differ by angle of $2\pi/N$) is well-defined. These choices
lead to the discrete target space $\Z_M$, i.e. $M$ points on a
circle. The coefficient group of the dual model is $\Z_M$ while the
dual target space is $\Z_N$. Thus under duality the number of points
in the target space is interchanged with the number of points in the
coefficient group of the cohomology. This is a novel feature of these
models. In addition to this interchanging, the radius of the target
space also undergoes a transformation.  Rather than including the
radius in the defining group, it appears in the Boltzmann weights. We
illustrate the relation of the Boltzmann weights and its character
coefficients for the case of a direct product of discrete groups: $\G
= \Z_{P_1} \oplus
\dots \oplus \Z_{P_n}$ ($P_i=N_iM_i$) and $\H =
\Z_{N_1} \oplus \dots \oplus \Z_{N_n}$, \bea B((g_1,\dots,g_n)) &=&
\left(\prod_{a=1}^n\sum_{m_a\in \Z }\right) \exp \left\{ -
G^{ij}\left(\frac{g_i}{P_i} + m_i \right) \left(\frac{g_j}{P_j}
+ m_j \right) \right\} \nn\\ 
b((\g_1,\dots,\g_n)) &=&
\sqrt{\frac{\pi}{{\widetilde G}}} \left(\prod_{a=1}^n \sum_{\m_a\in \Z }\right)
\exp \left\{ - \frac14({\widetilde G}^{-1})^{ij}
\left(\frac{\g_i}{P_i} + \m_i \right)\left(\frac{\g_j}{P_j} + \m_j \right) 
\right\} \nn \eea
where ${\widetilde G}^{ij} = G^{ij}/P_iP_j$ is the normalized
``metric'' on the original target space.  This demonstrates that even
in these models an inversion of the ``metric'' occurs, much like in
the toroidal compactifications, as one would expect.  For the model to
be self-dual the number of points in both the original and dual theory
should be identical. This is achieved if $N_i=M_i$ so that $P_i$ is a
perfect square.  Notice that in the limit of very large $N_i$ there
are a large number of points on the target space while the number of
elements in the spin group is of order $N_i^2$. One can then roughly view
the limit of infinite $N_i$ as the case where the target space becomes
a continuous circle while the spin variable reduces to the reals, thus
recovering the $S^1$ case.

Of course it is possible to choose $\G$ to be products of $\Z_{N_i^2}$
and $\R$, and $\H$ to be products of $\Z$ and $\Z_{N_i}$ to obtain
mixed target spaces which are explicitly self-dual. Some other choices
for $\G$ and $\H$ which are interesting on there own, but are not
self-dual, are $\G=U(1)$ and $\H=\Z_N$. With such target spaces the
string is allowed to wrap around the space only a finite number of
times before it is homotopically equivalent to zero windings.  In this
case, the dual target space is a discrete space, $\Z_N$, even though
the original target space was continuous. This nicely demonstrates
that even for Abelian isometries the dual need not have the same
fundamental group as the original target space (for the non-Abelian case see 
\cite{AlAlGBaLo93}).

It is possible to generalize these results to the case where the
lattice $\Sigma$ is a triangulation of an arbitrary
$(p+1)$-dimensional orientable manifold. There is one technical
constraint on $\Sigma$: $H^k(\Sigma,\Z)$ must be free Abelian, this is
automatic for the case of two-dimensional orientable manifolds but not
for higher dimensional spaces. The models in this case are written as,
$$ Z = \sum_{h\in H^k(\Sigma, \H)}
\;\sum_{\sigma\in C_{k-1}(\Sigma,\G)}\;
\prod_{l=1}^{\N_{k}} B\left( \left\langle \left(\delta\sigma + h
\right) , c_{k}^{(l)} \right\rangle\right)
$$ These are models in which a $\G/\H$-valued ``spin'' lives on the
$(k-1)$-dimensional cells of a $(p+1)$-dimensional lattice. For
example, if $k=2$ this describes a gauge theory on a $(p+1)$-dimensional
world-volume.  The duality transformations can be applied to this
model with very little effort. Simply replace the $0$-dimensional
objects with $(k-1)$-dimensional ones and $1$-dimensional objects with
$k$-dimensional ones. The dual model is a trivial extension of the
previous dual model, $$ Z =
\sum_{\h\in H^{p+1-k}(\Sigma, \G^*/\H^*)}
\;\sum_{\tsigma\in C_{p-k}(\Sigma,\G^*)}\;
\prod_{l=1}^{\N_{p+1-k}^*} b\left( \left\langle \left(\delta\tsigma + \h
\right) , c_{p+1-k}^{*(l)} \right\rangle\right)
$$ Self-dual models exist only when the previous relations among the
groups are satisfied and $p+1 = 2k$. Clearly $p=1$, $k=1$ is among
those and reproduces the string-theory case. The next case is $k=2$
and $p=3$. This is a gauge theory, with gauge group $\G$ defined on
the $4$-dimensional world-volume $\Sigma$ and the sum over $\H$-valued
cohomology is akin to summing over the topological sectors of the
theory. A continuum example is given by, $$ Z = \sum_{h\in H^2(\Sigma,
\Z)} \int {\cal D} X
\exp \left\{ - g^2 \int ( d X + 2\pi h) \wedge * (d X +2 \pi h) \right\}
$$ where the field $X$ is a real-valued one-form on $\Sigma$.  There
are of course many higher-dimensional analogues of such self-dual
models.

The author would like to thank L. D. Paniak for helpful discussions
and, along with J. Neilson, for commenting on the manuscript. I would
also like to thank G. W. Semenoff, K. Zarembo and A. Zhitnitsky for
useful input and the Niels Bohr Institute for its hospitality where
part of this work was completed.


\begin{thebibliography}{1234567} \newcommand{\bibi}[1]{\bibitem{#1}}
%---------------------------------------------------------
\newcommand{\authors}[1]{#1, } \newcommand{\journal}[1]{#1}
\newcommand{\volume}[1]{{\bf #1}} \newcommand{\myyear}[1]{(#1)}
\newcommand{\page}[1]{#1} \newcommand{\mytitle}[1]{}
\newcommand{\keywords}[1]{} \newcommand{\kw}[1]{}
%---------------------------------------------------------
\bibi{Tdual} \authors{K. Kikkawa and M. Yamasaki}
\journal{Phys. Lett.} \volume{B149} \myyear{1984} \page{357}.

\bibi{MoOv89} \authors{J. Molera and B. Ovrut} \journal{Phys. Rev.}
\volume{D40} \myyear{1989} \page{1146}.

\bibi{Du90}
\authors{M. J. Duff} \journal{Nucl. Phys.} \volume{B335} \myyear{1990} 
\page{610}.

\bibi{Si96} \authors{W. Siegel}
\journal{Phys. Rev.} \volume{D54} \myyear{1996} \page{2797}.

\bibi{GrKl90} \authors{D. Gross and I. Klebanov} \journal{Nucl. Phys.}
\volume{B344} \myyear{1990} \page{475}.  

\bibi{KrWa41}
\authors{H. A. Kramers and G. H. Wannier} \journal{Phys. Rev.}
\volume{60} \myyear{1941} \page{252}.  



\bibi{ShVa88} \authors{M. A. Shifman and A. I. Vainshtein}
\journal{Nucl. Phys.} \volume{B296} \myyear{1988} \page{445}.

\bibi{KoSh97} \authors{A. Kovner and M. A. Shifman}
\journal{Phys. Rev.} \volume{D56} \myyear{1997} \page{2396}.

\bibi{HaZh97} \authors{I. Halpern and A. Zhitnitsky}
\journal{hep-ph/9711398}.  

\bibi{Sc81} \authors{L. S. Schulman}
\journal{Techniques and Applications of Path Integration, John Wiley
\& Sons, 1981}.  

\bibi{AlAlGLo95} 
\authors{E. Alvarez, L. Alvarez-Gaume and Y. Lozano} 
\journal{Nucl. Phys. Proc. Suppl.} \volume{41} \myyear{1995}~\page{1}.

\bibi{GiPoRa94}
\authors{A. Giveon, M. Porrati and E. Rabinovici}
\journal{Phys. Rept.} \volume{244} \myyear{1994} \page{77}.

\bibi{Sp66}
\authors{E. H. Spanier}
\journal{Algebraic Topology, Springer-Verlag, 1966}.

\bibi{GaJaSe98} 
\authors{C. R. Gattringer, S. Jaimungal, G. W. Semenoff} 
\journal{hep-th/9712173 to appear in Phys.~Lett.~B}. 

\bibi{Bu87}
\authors{T. H. Buscher} \journal{Phys. Lett.} \volume{B194}
\myyear{1987} \page{59}.  

\bibi{AlAlGBaLo93}
\authors{E. Alvarez, L. Alvarez-Gaume, J. L. F. Barbon, Y. Lozano}
\journal{Nucl. Phys.} \volume{B415} \myyear{1994}~\page{71}.

\end{thebibliography}
 \end{document}